\documentclass[aps,prd,floats,nofootinbib,preprintnumbers]{revtex4}

\usepackage[utf8]{inputenc}
\usepackage{color,graphicx}
\usepackage{mathrsfs}
\usepackage{amssymb}
\usepackage{amsmath}
\usepackage{feynmp}
\usepackage{slashed}
\usepackage{latexsym}
\usepackage{verbatim}
\usepackage[normalem]{ulem}

\newcommand{\be}{\begin{equation}}
\newcommand{\ee}{\end{equation}}
\newcommand{\ba}{\begin{array}}
\newcommand{\ea}{\end{array}}
\newcommand{\bea}{\begin{eqnarray}}
\newcommand{\eea}{\end{eqnarray}}
\newcommand{\balg}{\begin{align}}
\newcommand{\ealg}{\end{align}}
\newcommand{\bit}{\begin{itemize}}
\newcommand{\eit}{\end{itemize}}
\newcommand{\trm}[1]{\textrm{#1}}
\newcommand{\mbf}[1]{\mathbf{#1}}
\newcommand{\tbf}[1]{\textbf{#1}}
\newcommand{\mcl}[1]{\mathcal{#1}}

\newcommand{\msc}[1]{\mathscr{#1}}

\newcommand{\tcr}[1]{\textcolor{red}{#1}}

\newcommand{\Mpc}{\trm{\Mpc}}
\newcommand{\yr}{\trm{\yr}}
\newcommand{\eV}{\trm{\eV}}

\newcommand{\nn}{\nonumber}

\newcommand{\vtw}{\vspace{.2cm}}

\begin{document}

\title{Imperfect Mirrors of the Standard Model}

\preprint{NUHEP-TH/16-02}

\author{Jeffrey M. Berryman, Andr\'e de Gouv\^ea, Daniel Hern\'andez, and Kevin J. Kelly}
\affiliation{Northwestern University, Department of Physics \& Astronomy, 2145 Sheridan Road, Evanston, IL~60208, USA}

\begin{abstract}

  Inspired by the Standard Model of particle physics, we discuss a mechanism for constructing chiral, anomaly-free gauge theories. 
The gauge symmetries and particle content of such theories are identified using subgroups and complex representations of simple anomaly-free Lie groups, such as $SO(10)$ or $E_6$. 
We explore, using mostly $SO(10)$ and the $\mathbf{16}$ representation, several of these ``imperfect copies'' of the Standard Model, including $U(1)^N$ theories, $SU(5)\otimes U(1)$ theories, 
$SU(4)\otimes U(1)^2$ theories with 4-plets and 6-plets, and chiral $SU(3)\otimes SU(2)\otimes U(1)$. A few  general properties of such theories are discussed, as well as how they might shed light on nonzero neutrino masses, the dark matter puzzle, and other phenomenologically relevant questions.

\end{abstract}

\maketitle

\setcounter{equation}{0}
\setcounter{footnote}{0}


\section{Introduction}

The Standard Model of particle physics (SM) is a chiral $SU(3)\otimes SU(2)\otimes U(1)$ gauge theory. Considering all fermion fields in the theory as, for example, left-handed Weyl spinors, this means that (i) all fermions are charged under the gauge symmetry, and (ii) there are no fermions with equal-and-opposite quantum numbers. These imply that fermion mass terms are forbidden by gauge invariance and all nonzero fermion masses are consequences of gauge symmetry breaking. A very attractive feature of the SM is all masses are proportional to same, unique source of mass, the scale of electroweak symmetry breaking.

Chiral gauge theories are, in general, anomalous. In the SM, quantum numbers are such that the contributions of the chiral fermions to the different anomalies cancel out exactly. This minor miracle implies that -- per generation -- all SM fermions are required in the sense that if any of them were absent, then the theory would be mathematically inconsistent. Considerations of anomaly cancellations contributed to the hypothesized existence of the bottom and top quarks, and of the tau-type neutrino, years before direct discovery. The peculiar quantum numbers of the fermion fields, required by anomaly cancellations, also imply that the SM Lagrangian is invariant under accidental global symmetries, including baryon and lepton numbers. 

The SM is known to be incomplete. The dark matter puzzle and nonzero neutrino masses imply that degrees of freedom outside of the SM exist. It is interesting to take the SM Lagrangian as guidance and investigate whether the new degrees of freedom are also chiral fermions, charged under a gauge symmetry. Identifying possible chiral gauge theories is a nontrivial exercise, mostly because of the anomaly cancellation requirement~\cite{BatraDobrescu, BabuSeidl}. 

Here, we discuss a novel yet powerful mechanism for generating chiral, anomaly-free gauge theories. We refer to these models as \emph{imperfect mirrors} of the SM. In a nutshell, we start with simple gauge theories that are anomaly-free and admit complex representations -- these include $SO(10), SO(14)$, and $E_6$ -- and identify gauge symmetries with subgroups of the theory and the fermion fields with representations of the simple gauge group. The mechanism is discussed in detail and is easy to implement. We explore, using mostly $SO(10)$ and the $\mathbf{16}$ representation, several chiral, anomaly-free gauge theories, including quasi-perfect copies of the SM,  and discuss qualitatively general properties of such theories and how they might shed light on the the neutrino mass and dark matter puzzles.

%
%
%
%
%

\setcounter{equation}{0}
\section{Chiral $U(1)$ theories from simple orthogonal groups}
\label{sec:U1}


\newcommand{\Gs}{\msc{G}}
In Ref.~\cite{firstpaper}, a general method was described to obtain anomaly-free abelian theories with chiral particle content. Let us briefly review those results.
The starting point is a gauge field theory based on a simple, anomaly-free Lie group $\msc{G}$. Such groups were identified by Georgi and Glashow in Ref.~\cite{GGnoanomalies}. The complete list is:
\begin{enumerate}
\item $SU(2)$.
\item The infinite sequence $Sp(2N)$.
\item The infinite sequences $SO(2N+1)$ and $SO(2N+2)$ with the exception of $SO(6)$.
\item The 5 exceptional groups: $G_2$, $F_4$, $E_6$, $E_7$ and $E_8$.
\end{enumerate}

Consider an anomaly-free simple group $\Gs$ and a semisimple group $\Gs' \subset \Gs$ which is not necessarily anomaly-free. Starting from a representation $R$ of $\Gs$, one can construct anomaly-free theories based on the gauge group $\Gs'$ by decomposing $R$ into a direct sum of representations $R'_i$ of $\Gs'$, and introducing the corresponding fermionic content. Such a procedure, however, may lead to nonchiral $\Gs'$ theories, i.e., to trivial cancellation of anomalies. Focusing on the abelian case, $\Gs' \equiv U(1)$, in order to end up with a chiral $\Gs'$ gauge theory, it is necessary for the initial $\Gs$ representation $R$ to be a complex representation~\cite{firstpaper}.

Of the anomaly-free groups listed above, only those of the form $SO(4N+2)$ and the exceptional group $E_6$ have complex representations. The group with lowest dimensionality of these is $SO(10)$, a well-known unification group~\cite{Georgi:1974my,FMSO(10)}. For any $SO(4N+2)$ algebra, the smallest complex representation is the fundamental spinor of dimension $2^{2N}$. Hence, the simplest representation that satisfies all requirements is the $\mbf{16}$ of $SO(10)$. Most of this work thus focuses on $\Gs\equiv SO(10)$ and $R \equiv \mbf{16}$, although we will mention other possibilities in Section~\ref{sec:extensions}.

Without loss of generality, we can identify $\msc{G}'$ with a $U(1)$ subgroup generated by an element $H \in\mcl{H}$ of the Cartan subalgebra of $\msc{G}$. The charges of the fermionic matter content with respect to $\msc{G}'$ are then given by the eigenvalues of $H$. 
We denote the $\Gs'$ charges by $q_i$ with $i = 1,\dots,d$, where $d$ is the dimension of the $\Gs$ representation $R$. Note that for compact $\Gs$, $H$ can be assumed Hermitian which implies that the $q_i$ can be taken to be real.

A generic element of the Cartan subalgebra in the $\mbf{16}$ representation of $SO(10)$ has the form:
\begin{align}
H(a,b,c,d,e) = \frac{1}{N}\trm{diag}\{ & a + b + c + d + e, \, -a + b + c + d - e,\, a - b + c + d - e, \, -a - b + c + d + e,  \nn \\
&  a + b - c + d - e, \, -a + b - c + d + e,\, a - b - c + d + e, \, -a - b - c + d - e, \nn \\
& a + b + c - d - e,\, -a + b + c - d + e,\, a - b + c - d + e,\, -a - b + c - d - e, \nn \\
& a + b - c - d + e, \, -a + b - c - d - e,\, a - b - c - d - e,\, -a - b - c - d + e \}   \,, \label{eig-list}
\end{align}
where $a$, $b$, $c$, $d$ and $e$ are arbitrary real numbers and $N$ is a normalization factor. These eigenvalues form the list of $\Gs'$ charges $\{q_i\}$. For generic values of $a,b,c,d,e$, for any $q \in \{q_i\}$, $-q \notin \{q_i\}$. One can quickly check that for any $a,b,c,d,e$, the $\{q_i\}$ are solutions to the anomaly $\Gs'\equiv U(1)$ equations,
\be
\sum_{i=1}^{16} q_i = 0\,,\quad \quad \sum_{i=1}^{16} q_i^3 = 0 \label{anomaly-eqs} \,.
\ee



An anomaly-free $\Gs'$ model is specified by a choice of $\{a,b,c,d,e\}$. Some properties of  Eq.~\eqref{eig-list} are of note. Chirality is destroyed if at least one of $a$, $b$, $c$, $d$ or $e$ vanishes. In that case, for every $q$ in Eq.~\eqref{eig-list}, $-q$ is also present. Hence, in order to obtain a chiral theory, it is necessary to take all of $a$, $b$, $c$, $d$ and $e$ different from zero. 
%
%

\newcommand{\abcde}{\{a, b, c, d, e\}}
Consider next the effect of flipping the sign of one of $\{a, b, c, d, e\}$. It holds that:
\be
H(-a,b,c,d,e) = H(a,-b,c,d,e) = H(a,b,-c,d,e) = H(a,b,c,-d,e) = H(a,b,c,d,-e) = - H(a,b,c,d,e) \,.
\label{eq:positive}
\ee
Flipping the sign of one among $\{a, b, c, d, e\}$ is equivalent to a model where all the charges have been reversed in sign. Since the overall sign of the charges is a convention, this implies that there is no loss of generality in considering all of $a$, $b$, $c$, $d$, $e$ strictly positive. This has an immediate corollary:
%
\emph{The largest $\Gs'$ charge is $a+b+c+d+e$ and it appears only once, the absolute values of all other charges being necessarily smaller.}
Hence, in every anomaly-free model derived following this method, there can be only one state with highest gauge abelian charge.

This is particularly important since, in the following section, we will extend the procedure to the case in which $\Gs'$ has nonabelian components in addition to the $U(1)$ symmetry. The particle content in $R$ will organize into multiplets of the nonabelian symmetry, and all states in a multiplet must have the same abelian charge. Hence, the state with highest $U(1)$ charge must transform as a singlet under any nonabelian piece of $\Gs'$. As an example, consider the SM. The state with highest hypercharge (the $U(1)$ inside $\Gs'$ in that case) is the left-handed antielectron $e^c$. 
This analysis implies that the $e^c$ must be a singlet of both color $SU(3)$ and weak $SU(2)$, as it is indeed the case.

It is useful to introduce a geometric picture of this procedure. The Cartan subalgebra of a group forms a vector space, which in the case of $SO(10)$ is five-dimensional. Taking into account that the absolute scale of charges is a matter of convention, choosing specific values for $a$, $b$, $c$, $d$ and $e$ amounts to selecting a one-dimensional subspace, i.e., a ray, in $\abcde$ space. Nonchiral models are located within hyperplanes where at least one coordinate vanishes, while chiral theories live inside a hyperquadrant. Sections of $\{a,b,c,d,e\}$ space are shown in Fig.~\ref{fig1}.


Next, we restrict discussion to theories with integer charges. From this requirement, either none, two, or four out of $a$, $b$, $c$, $d$ and $e$ can be half-integers, while the rest are integers. Example theories include: 
\begin{itemize}
\item $a=b=c=d=e$. This case comprises one model up to charge rescalings, that is a ray in $\abcde$ space. For $a=1$, we obtain the following charges and multiplicities:
  \begin{center}
    \begin{tabular}{c|c|c|c}
      Charge & $5$ & $-3$ & $1$ \\ \hline
      Multiplicity & 1 & 5 & 10
    \end{tabular}
  \end{center}
  We call this model the (5).
\item $a=b=c=d$, $e$. Here, different models can be specified by the ratio $r = e/a$ which labels the specific ray in the space of the Cartan subalgebra along which we intend to gauge. For instance, the following model  is achieved with $a =b=c=d=1/2$, $e=1$, corresponding to $r=2$:
  \begin{center}
    \begin{tabular}{c|c|c|c|c|c}
      Charge & $3$ & $-2$ & $1$ & $0$ & $-1$ \\ \hline
      Multiplicity & 1 & 4 & 6 & 4 & 1
    \end{tabular}
  \end{center}
  We call any model of this type, a (4, 1).
  Note that as long as we are building purely $U(1)$ models, neutral states as well as pairs of opposite charges give a vanishing contribution to the anomaly equations, Eq.~\eqref{anomaly-eqs}, and can be removed from the list of states. From this point of view, the model above has ten chiral fermions. 
\item $a=b=c$, $d=e$. Again, we define $r = e/a$. For $a=b=c=1$, $d=e=1/2$ ($r=1/2$), we find:
  \begin{center}
    \begin{tabular}{c|c|c|c|c|c|c}
      Charge & $4$ & $-3$ & $-2$ & $2$ & $0$ & $1$ \\ \hline
      Multiplicity & 1 & 2 & 3 & 1 & 3 & 6
    \end{tabular}
  \end{center}
  We call any model of this type, a (3, 2). The model above contains eleven chiral fermions. 
\end{itemize}
More generally, we label models where the set $\abcde$ can be split into subsets of $j_1$, $\dots$, $j_k$ equal numbers, $(j_1,\dots, j_k)$ models. Clearly, for the \tbf{16} of $SO(10)$, $\sum j_i = 5$, that is, the $j_i$s constitute a partition of 5.

The alert reader will have noticed that the charges in these examples organize suggestively. In particular, in the $(3, 2)$ case the charges seem to organize into multiplets resembling the SM particle content. A bit of experimentation shows that in all these models, although the charges themselves depend on $r$, the number of particles with identical charges does not. In the (5) case, for example, the charges mimic the multiplets in the $SU(5)$ Georgi-Glashow unification model~\cite{GGSU(5)}. We prove next that this resemblance is not accidental.


\setcounter{equation}{0}
\section{Nonabelian symmetries}
\label{sec:nonabelian}



The $\{a,b,c,d,e\}$ formalism can also be used to identify gauge theories based on nonabelian groups. The key fact is the following result:
\emph{Any $(j_1,\dots,j_k)$ model accepts the simultaneous introduction of nonabelian, anomaly-free $SU(j_i)$ gauge groups for any $j_i \geq 2$.}


We begin by illustrating the result for the case of $SU(2)$. Without losing generality, we show that there is an $SU(2)$ subalgebra that commutes with $H(a,a,\dots)$. The proof proceeds by construction. The $2^N$-dimensional spinor representation of $SO(2N+2)$ acts on a vector space that is conveniently represented as the tensor product of $N$ 2-dimensional vector spaces~\cite{georgibook}. We choose a basis in this space with its elements given by:
\be
\label{binary-states}
|k_1\rangle|k_2\rangle \dots |k_N\rangle
\ee
where $k_i = \pm 1$.

Generators of $SO(2N+2)$ can be written as tensor product operators composed of Pauli matrices acting independently on the 2-dimensional vector subspaces. We define a basis of generators as follows:
\begin{align}
  H^s & = \sigma_3^{s} \,, \quad \trm{for } s = 1,\dots,N, \\
  H^5 & = \sigma_3^{1}\otimes\sigma_3^{2}\otimes \dots\otimes \sigma_3^{N}, \\
  A_i^s & = \bigotimes_{r=1}^{s-1}\sigma_3^r \otimes \sigma_i^s  \,, \quad \trm{for } s = 1,\dots,N, \\
  B_i^s & = \sigma_i^s \otimes \bigotimes_{r=s+1}^{N}\sigma_3^r  \,, \quad \trm{for } s = 1,\dots,N, \\
  M_{ij}^{st} & = \sigma_i^s\otimes \bigotimes_{r = s+1}^{t-1}\sigma_3^r\otimes \sigma_j^t, \quad  \trm{for } s,t = 1,\dots,N,\; s\neq t,
\end{align}
where $\sigma_i^{r}$ acts on the $r$th vector subspace. The goal is to prove that there is an $SU(2)$ subalgebra that commutes with the generator
\begin{equation}
H = \sum_{a=1}^5 \beta_a H^a,
\end{equation}
if $\beta_1 = \beta_2$.


Without loss of generality, we define the following:
\begin{align}
T_1 = \frac{1}{4}\left(M_{11}^{12} + M_{22}^{12}\right)\,,\quad T_2 = \frac{1}{4}\left(M_{12}^{12} - M_{21}^{12}\right)\,,\quad T_3 = \frac{1}{4}\left(H^2 - H^1\right).
\end{align}
The commutators among these are
\begin{align}
[T_1, H] &= \frac{i}{2}\left(\beta_1 - \beta_2\right) \left(M_{12}^{12} - M_{21}^{12}\right) = 2i\left(\beta_1 - \beta_2\right) T_2,\\
[T_2, H] &= \frac{i}{2}\left(\beta_1 - \beta_2\right) \left(M_{11}^{12} + M_{22}^{12}\right) = 2i\left(\beta_1 - \beta_2\right) T_1,\\
[T_3, H] &= 0,\\
[T_1, T_2] &= \frac{i}{4}\left(H^2 - H^1\right) = iT_3.
\end{align}
The generators $T_{1,2,3}$ form an $SU(2)$ subalgebra belonging to $SO(10)$ that commutes with $H$ if $\beta_1 = \beta_2$ (regardless of $\beta_3$, $\beta_4$, $\dots$), which is what we wanted to prove.

In order to extend the proof to an $SU(K)$ subgroup, we need to construct the corresponding subalgebra for $\alpha^1 = \alpha^2 = \dots = \alpha^K$. Such a subalgebra can be defined by the following choice of generators: 
\begin{align}
  F^n & = \frac{1}{\sqrt{2^Nn(n+1)}} \left[ \left(\sum_{i=1}^nH^i\right) - nH^{n+1} \right] \,,\quad n = 1,\dots,K-1\,, \\
  E^{lm} & = \left\{ \begin{array}{l}
    \frac{1}{\sqrt{2^{N+1}}} (M_{11}^{lm} + M_{22}^{lm})\,,\quad \trm{for } l,m=1,\dots,K ;\; l < m \,, \\
    \frac{1}{\sqrt{2^{N+1}}} (A_1^l - B_1^l) \,,\quad \trm{for } l=1,\dots,K ;\;  m = K+1 \,,
  \end{array} \right.  \\
  \overline{E}^{lm} & = \left\{ \begin{array}{l}
    \frac{1}{\sqrt{2^{N+1}}} (M_{12}^{lm} - M_{21}^{lm})\,,\quad \trm{for } l,m=1,\dots,K ;\; l < m \,, \\
    \frac{1}{\sqrt{2^{N+1}}} (A_2^l - B_2^l) \,,\quad \trm{for } l=1,\dots,K;\;  m = K+1 \,.
  \end{array} \right. 
\end{align}
It is straightforward to check that this choice of generators indeed satisfies the $SU(K)$ commutation relations.

\begin{figure}
\begin{center}
\begin{minipage}[t]{17cm}
  \includegraphics[width=4.4cm]{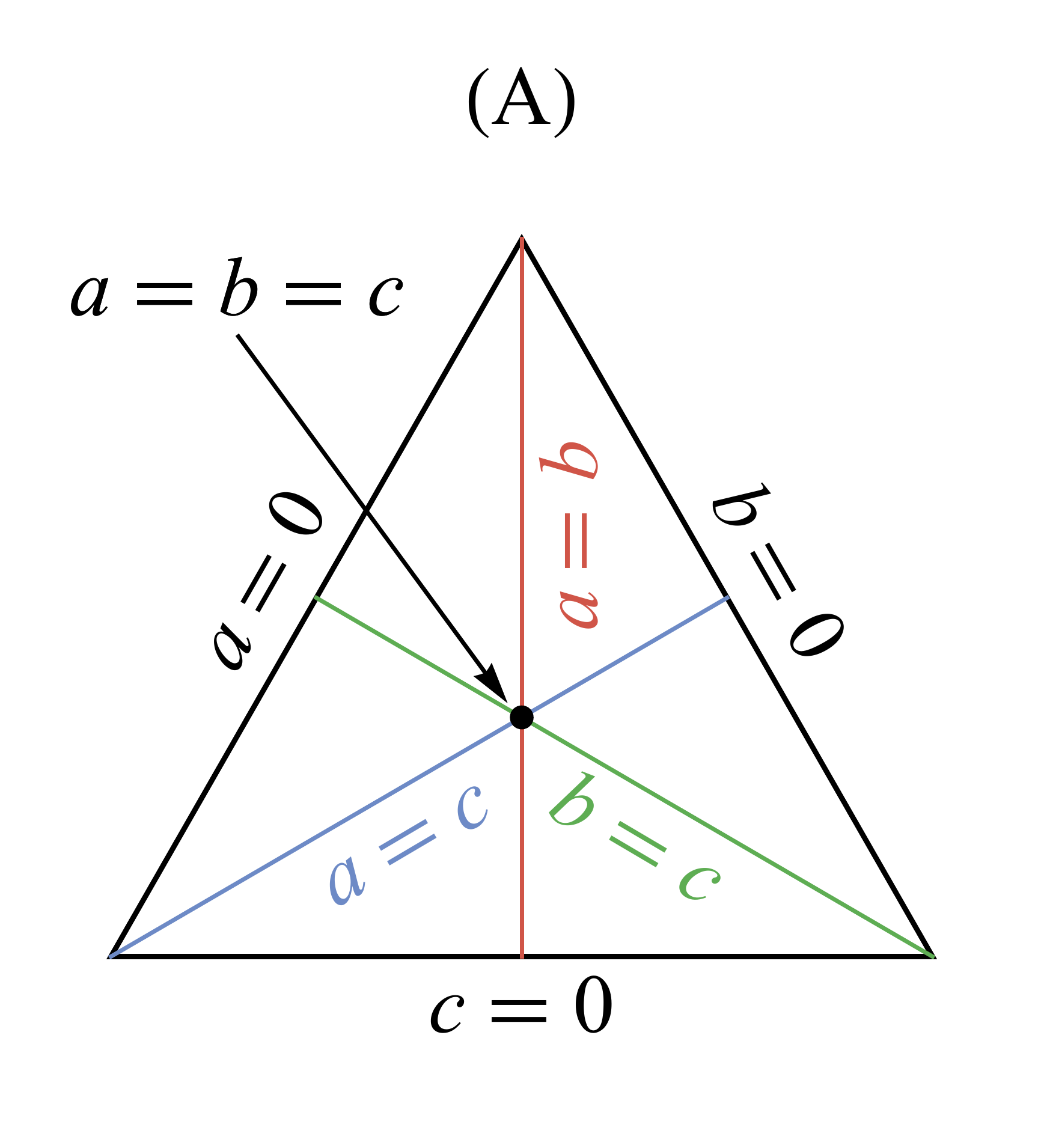}
  \includegraphics[width=6.0cm]{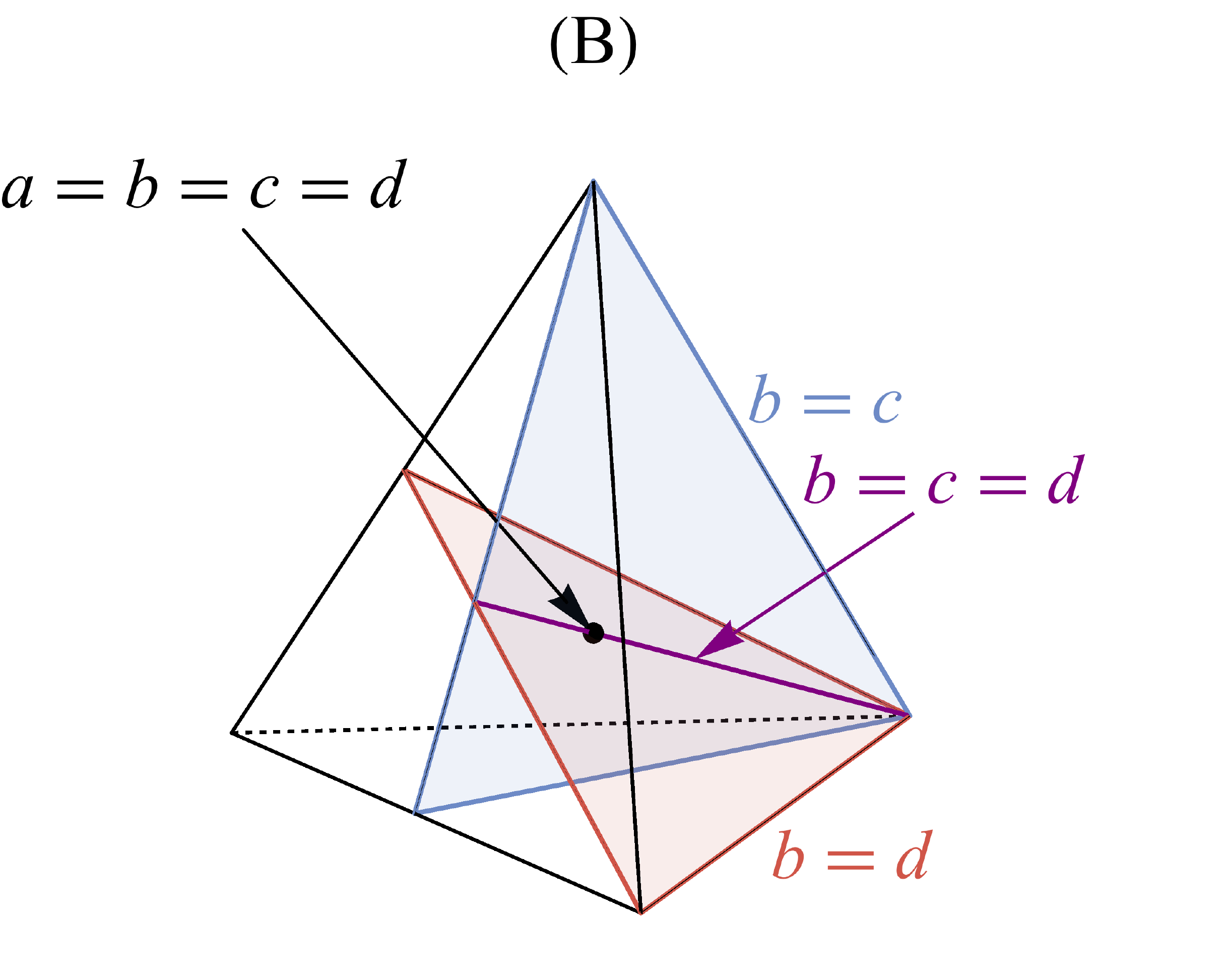}
  \includegraphics[width=6.1cm]{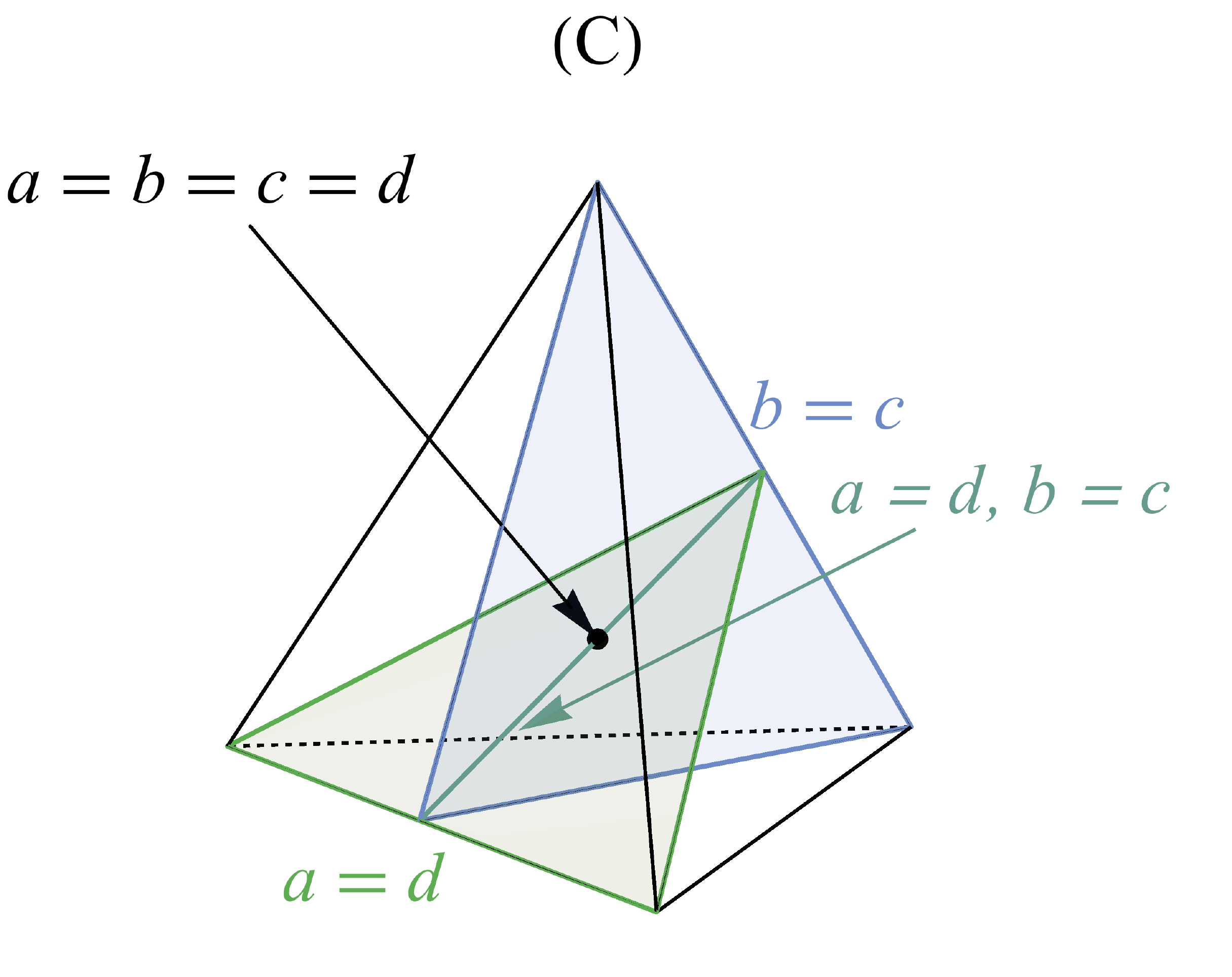}
\end{minipage}
\caption{Cross sections of the $\{a,b,c,d,e\}$ vector space. Figure (A) represents the plane $a+b+c = 3d$,  $d=e$ with $a,b,c>0$. The centroid of the triangle corresponds to the $SU(5)$ model, the medians correspond to the $SU(2)\otimes SU(2)$ models ($d = e$, $a = b$). Nonchiral models live in the edges of the triangle where at least one of $a$, $b$ or $c$ vanishes. Figures (B) and (C) represent the tetrahedron $a+b+c+d=4e$ for positive $a,b,c,d$. The lines joining (B) the vertex with the opposite face of the tetrahedron and (C) the midpoints of two opposing edges have been drawn. They represent models with larger symmetry.}
\label{fig1}
\end{center}
\end{figure}

As we have stressed before, the numbers $\{a,b,c,d,e\}$ span a vector space in which each model corresponds to a ray. The set of models corresponds therefore to 4D projective space. A graphical representation of some features of this space is depicted in Fig.~\ref{fig1}. In Fig.~\ref{fig1}(A), a planar cross section of 4D projective space in the coordinates $\{a, b, c\}$ is shown. The conditions selecting this hyperplane are
\be
d=e\,,\quad a+b+c = 3d. \label{conds1}
\ee
The restriction to positive $a,b,c,d,e$ extracts an equilateral triangle out of this plane whose centroid, the point $a=b=c=d=e$, has maximal $SU(5)$ symmetry. The medians of the triangle also have enhanced symmetry, in this case $SU(2)\otimes SU(2)\otimes U(1)$. Note that the conditions in Eq.~\eqref{conds1} are invariant under simultaneous rescalings of all coordinates as befits a projective space.

A solid cross section is shown in Fig.~\ref{fig1}(B) and Fig.~\ref{fig1}(C) corresponding the requirement $a+b+c+d = 4e$. The barycenter of the tetrahedron is again the maximally symmetric point $a=b=c=d=e$, and the segments depicted, dropping from a vertex to the centroid of the opposite face or joining the midpoints of two opposite edges, have again enhanced symmetry, $SU(3)$ and $SU(2)\otimes SU(2)$ respectively.

In summary, we have shown that chiral gauge theories based on the \tbf{16} representation of $SO(10)$ are in a one-to-one correspondence with a set of five strictly positive numbers. The substitution of these in Eq.~\eqref{eig-list} yields charges with respect to a chiral, anomaly-free $U(1)$ symmetry. If two or more among $\{a,b,c,d,e\}$ are equal, the model can also accommodate enhanced, nonabelian symmetries, as described above.



\setcounter{equation}{0}
\section{Applications}
\label{sec:applications}

The formalism developed above can be used to quickly explore the infinite space of chiral gauge theories. Here, we discuss several models, highlighting some of their main features and how they may be useful for addressing phenomenology questions. In this section, we concentrate on scenarios seeded by the $\mathbf{16}$ representation of $SO(10)$. We will briefly comment on higher representations of $SO(10)$, the $\mathbf{64}$ representation of $SO(14)$, and the $\mathbf{27}$ representation of $E_6$ in Sec.~\ref{sec:extensions}. 

Simple chiral $U(1)$ theories have been explored in, for example, Refs.~\cite{BatraDobrescu,firstpaper}, 
and it is possible to organize and classify them using different criteria. We can use the highest charge as an organizing principle. If we impose that all $U(1)$ charges are integers, the model characterized by $(a = b = c = d = 1/2, e = 1)\equiv \{\frac{1}{2},\frac{1}{2},\frac{1}{2},\frac{1}{2},1\}$ is such that the charge of the highest-charged fermion is as small as possible (and equal to 3). There is only one model with highest charge 3. The three models with highest charge 4 are $\{\frac{1}{2},\frac{1}{2},1,1,1\}$, $\{\frac{1}{2},\frac{1}{2},\frac{1}{2},1,\frac{3}{2}\}$,  and $\{\frac{1}{2},\frac{1}{2},\frac{1}{2},\frac{1}{2},2\}$. There are more models with highest charge 5, 6, etc. The number of models is roughly proportional to the combinatorial partitions of the highest charge.

One can also characterize $U(1)$ models by the number of chiral fermions.\footnote{Keep in mind that fermions with charge zero and pairs of fermions with opposite charge are not chiral and contribute trivially to the anomaly cancellation constraints.} Starting from the $\mathbf{16}$ representation of $SO(10)$, chiral $U(1)$ models will contain at most 16 chiral fermions. On the other hand, the smallest number of chiral fermions is 5, as can be demonstrated quite generally: theories with 1, 2, 3, and 4 chiral fermions cannot satisfy the anomaly cancellation conditions Eqs.~\eqref{anomaly-eqs}~\cite{BatraDobrescu,firstpaper}. 
The following solution of Eqs.~\eqref{anomaly-eqs} with highest charge $10$ requires only five fields: $q_1=2, q_2=4, q_3=-7, q_4=-9, q_5=10$.
This is the $\{\frac{1}{2},\frac{3}{2},2,\frac{5}{2},\frac{7}{2}\}$ model. 

The scalar sector of the theory, responsible for $U(1)$ symmetry breaking, also helps define the phenomenological consequences of the model. If none of the fermions acquire Majorana masses, models with an odd number of chiral fermions will contain one massless Weyl fermion.\footnote{Forbidding Majorana masses is equivalent to requiring the Yukawa sector to preserve a subset of accidental global symmetries.} This is the case of the SM with its minimal Higgs sector (i.e., only one Higgs doublet $H$). For each generation there are fifteen fermions, and there is one massless left-handed neutrino after spontaneous gauge symmetry breaking. The existence of a Higgs triplet $T$, such that the $LTL$ Yukawa interaction is allowed, violates lepton number, and $\langle T\rangle\neq 0$ would lead to neutrino Majorana masses. The scenario explored in Ref.~\cite{firstpaper} also contains an odd number of hidden-sector fermions (eleven), and one of them is massless after $U(1)_{\nu}$ symmetry breaking. The Yukawa sector of the theory is such that accidental global symmetries prevent this left-handed ``antineutrino'' from acquiring a Majorana mass~\cite{firstpaper}.  

\begin{figure}
\begin{center}
\includegraphics[width=0.5\linewidth]{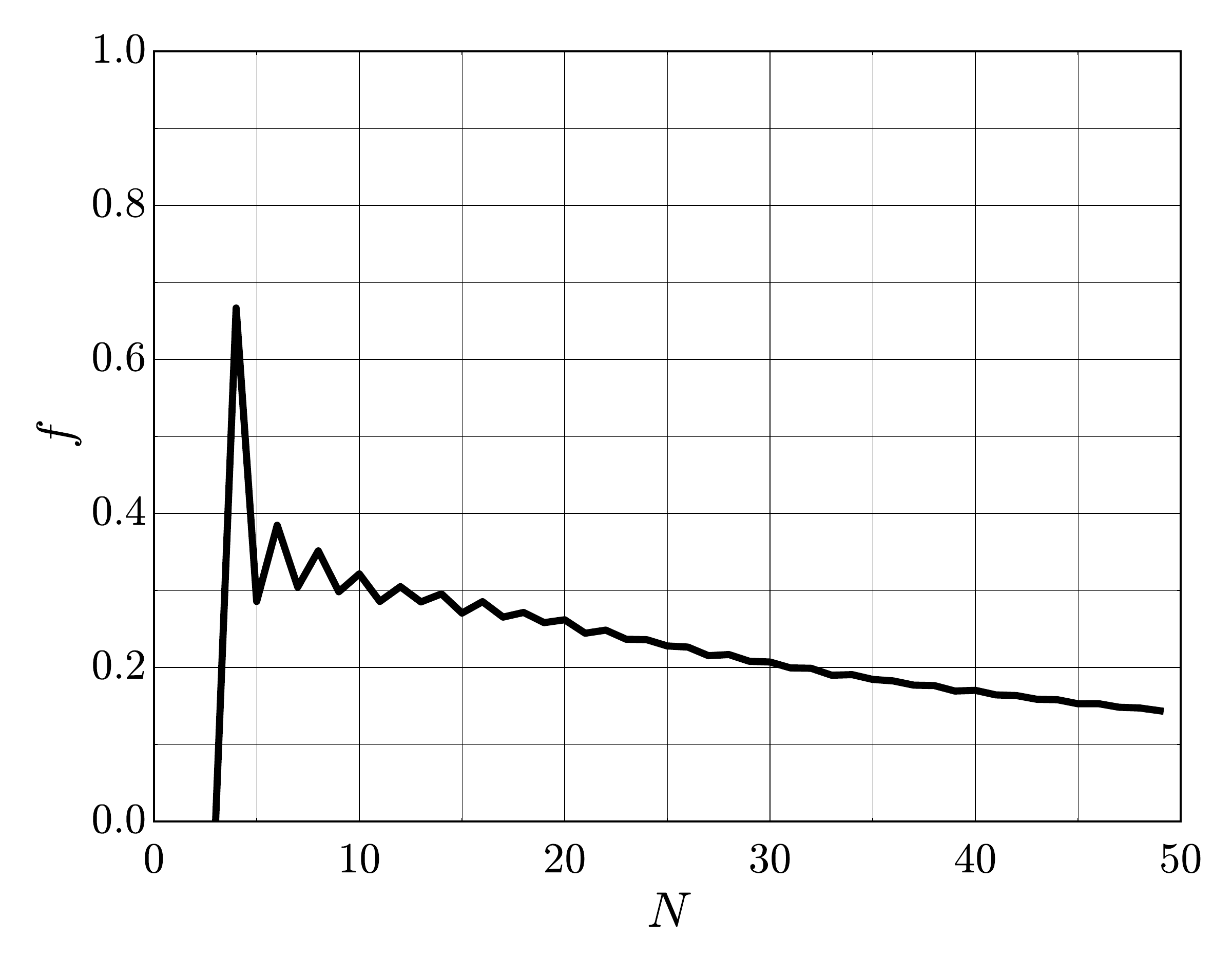}
\caption{Fraction $f$ of $U(1)$ models where all fermions, except at most one, obtain nonzero Dirac masses from the expectation value of one scalar field, as a function of the highest charge $N$.}
\label{fig2}
\end{center}
\end{figure}

For a generic chiral $U(1)$ model, one can compute the smallest number of scalar fields required to make sure that (i) at most one fermion is massless after spontaneous symmetry breaking, and (ii) all massive fields are Dirac fermions. In the SM, and in the model explored in Ref.~\cite{firstpaper}, this number is one. In the scenario with the smallest number of fermions ($\{\frac{1}{2},\frac{3}{2},2,\frac{5}{2},\frac{7}{2}\}$, above), two scalar fields are required. We investigated the fraction of models where Dirac masses for all fermions (except at most one) can come from the vacuum expectation value of only one scalar field. As a function of the highest charge, the fraction of scenarios with only one Higgs decreases, as depicted in Fig.~\ref{fig2}. Oscillations arise because these scenarios are more likely for an even highest charge.

The same procedure can be used to generate chiral $U(1)^N$ theories. In the $SO(10)$-seeded scenarios under consideration here, $N\le 5$. Models with $U(1)^2$ symmetry, for example, correspond to a pair of vectors $\{a_1,b_1,c_1,d_1,e_1\}$, $\{a_2,b_2,c_2,d_2,e_2\}$ that are linearly independent, while $U(1)^4$ models correspond to four linearly independent vectors in the $\{a,b,c,d,e\}$ space. In general, $U(1)^N$ models will include particles non-trivially charged under all $U(1)$ symmetries. Models with $N>5$ can also be constructed, but they require the use of other seed groups, including $E_6$, which can accommodate $U(1)^6$ chiral models and $SO(14)$, which can accommodate $U(1)^7$ chiral models. 

Turning to the nonabelian scenarios, anomaly-free $SU(N)$ theories have been investigated for instance in Ref.~\cite{Eichten}. In our formalism, the $(5)$ model, introduced in Sec.~\ref{sec:U1}, corresponds to $\{a,a,a,a,a\}$. As shown in Sec.~\ref{sec:nonabelian}, we can choose an $SU(5)$ subgroup of $SO(10)$ that commutes with the $U(1)$ subgroup characterized by $\{a,a,a,a,a\}$ such that the emerging particle content can be organized into a chiral, gauge-invariant $SU(5)\otimes U(1)$ theory. For $a=1$, $\{1,1,1,1,1\}$ translates into 
\be
\mathbf{10}_1 \,,\quad \mathbf{\overline{5}}_{-3}  \,,\quad \mathbf{1}_5 \,,
\label{model5}
\ee
where the number in bold-face refers to the $SU(5)$ representation and the subscript to the $U(1)$ charge. As emphasized before, there is an $SU(5)$ singlet, and the singlet has the highest $U(1)$ charge. In the $(5)$ scenario, the particle content and charge assignments are unique, modulo an overall (unphysical) rescaling of all $U(1)$ charges. The particle content also agrees with that of the $SO(10)$ grand-unified theory (GUT) where the GUT symmetry-breaking follows the path $SO(10)\to SU(5)\otimes U(1)$. The $U(1)$ in this case is sometimes referred to as $U(1)_X$, and is a unique linear combination of $U(1)_{B-L}$ and $U(1)_Y$, where $B-L$ refers to baryon number minus lepton number. 

While we used the $U(1)$ subgroup of $SO(10)$ to identify the chiral $SU(5)$ particle content, one is allowed to turn off the $U(1)$ symmetry. The leftover $SU(5)$ theory is still guaranteed to be anomaly-free. Investigating Eq.~\eqref{model5}, we can conclude that a chiral $SU(5)$ model with one $\mathbf{10}$ and one $\mathbf{\overline{5}}$ is anomaly free, as is well known. This is the only chiral $SU(5)$ model one can derive from the $\mathbf{16}$ of $SO(10)$.   

The $(4,1)$ models, introduced in Sec.~\ref{sec:U1}, can be interpreted as chiral $SU(4)\otimes U(1)$ models. In particular, the $U(1)$ is characterized by $\{a,a,a,a,e\}$, and translates into the following $SU(4)\otimes U(1)$ particle content, for $a=1,r=e/a$:
\be
\mathbf{6}_r \,,\quad \mathbf{4}_{2-r}  \,,\quad \mathbf{\overline{4}}_{-2-r} \,,\quad \mathbf{1}_{4+r} \,, \quad \mathbf{1}_{-4+r} \,.
\label{model6}
\ee
The number in bold-face refers to the $SU(4)$ representation and the subscript to the $U(1)$ charge. There are, in general, two singlets, and the highest $U(1)$ charge is $4+r$. Similar to the $SU(5)\otimes U(1)$ case, the $SU(4)$ particle content is unique. In this case, one can also choose to ignore the $U(1)$ part of the theory, and only explore the $SU(4)$ model with the particle content spelled out in Eq.~\eqref{model6}, ignoring the singlet fields. This model, unlike the $SU(5)$ model in Eq.~\eqref{model5}, is not chiral since there is both a $\mathbf{4}$ and a $\mathbf{\overline{4}}$, which can combine into a Dirac fermion, and a $\mathbf{6}$, which is a real representation and allows for a vector-like mass term. In summary, while the $SU(4)\otimes U(1)$ is a chiral model for any value of $r\neq 0$, the $SU(4)$ model is not. 

Given that the rank of $SO(10)$ is greater than that of $SU(4)\otimes U(1)$, it is possible to gauge a different $U(1)$ and consider chiral $SU(4)\otimes U(1)^2$ models. The two $U(1)$s are characterized by, for example, $\{1,1,1,1,r\}$ and $\{1,1,1,1,s\}$, $r\neq s$. Note that an analogous choice is to pick the second $U(1)$ to be nonchiral, e.g., $\{0,0,0,0,s'\}$. The reason is that $\{1,1,1,1,r\}$, $\{1,1,1,1,s\}$, and $\{0,0,0,0,s'\}$ are not linearly independent: $\{1,1,1,1,s\}+\{0,0,0,0,s'\}=\{1,1,1,1,r\}$, $r=s+s'$.  

One scalar field with a nonzero vacuum expectation value is sufficient to render all fermions in Eq.~\eqref{model6} massive. Explicitly, if we add to the theory a scalar 4-plet of $SU(4)$ with $U(1)$ charge $-2$, $\mathbf{4}^H_{-2}$, the following Yukawa Lagrangian preserves gauge invariance,
\be
{\cal L}_{\rm Yukawa} = \lambda_1 \mathbf{4}^H_{-2} \mathbf{6}_r\mathbf{4}_{2-r} + \lambda_2 \mathbf{4}^H_{-2} \mathbf{1}_{4+r}\mathbf{\overline{4}}_{-2-r} + \lambda_3 (\mathbf{4}^H_{-2})^{\dagger} \mathbf{6}_r \mathbf{\overline{4}}_{-2-r} + \lambda_4 (\mathbf{4}^H_{-2})^\dagger \mathbf{1}_{-4+r} \mathbf{4}_{2-r} + h.c.\,, 
\label{model6_y}
\ee
where the $(\mathbf{4}^H_{-2})^{\dagger}$ transforms like a $\mathbf{\overline{4}}$ with charge $+2$. The case $r=4$ is special. Here, one of the $SU(4)$ singlets has zero $U(1)$ charge and is allowed to have a Majorana mass, unrelated from the $SU(4)\otimes U(1)$ symmetry breaking scale. In the $r=4$ case, one may also consider a model with a reduced particle content, in which the gauge-singlet fermion is ignored. 

For illustrative purposes, we discuss in more detail the $\{\frac{1}{2},\frac{1}{2},\frac{1}{2},\frac{1}{2},2\}$ $SU(4)\otimes U(1)$ (corresponding to $r=4$ above, redefining all charges by a factor of 1/2) model without the gauge singlet fermion. The Yukawa Lagrangian is given by Eq.~\eqref{model6_y} with $\lambda_4\equiv0$, and the number of chiral fermions is odd. 
Following the symmetry-breaking pattern $SU(4)\otimes U(1)\to SU(3)\otimes U(1)'$, 14 of the 15 fields get Dirac masses, except for part of the $ \mathbf{4}_{-1}$, which remains massless. The degrees of freedom reorganize into triplets, anti-triplets and singlets of $SU(3)$, as follows:
\be
\mathbf{6}_2 \to \mathbf{\overline{3}}_{1/3}\oplus \mathbf{3}_{2/3} \,,\quad \mathbf{4}_{-1} \to \mathbf{3}_{-1/3}\oplus \mathbf{1}_{0}  \,,\quad \mathbf{\overline{4}}_{-3}\to \mathbf{\overline{3}}_{-2/3}\oplus \mathbf{1}_{-1} \,,\quad \mathbf{1}_{4} \to \mathbf{1}_{1}\,,
\label{model666}
\ee
while $\mathbf{4}^H_{-1}\to \mathbf{3}^H_{-1/3} \oplus \mathbf{1}^H_0$, including the degrees of freedom eaten by the heavy vector bosons. Here, the subscripts indicate the corresponding $U(1)'$ charges. The broken generators (7 of them) manifest themselves as massive gauge bosons $W_3$, which are $\mathbf{3}_{-1/3}$, and $Z_3$, which is an $SU(3)\otimes U(1)'$ singlet. From the $SU(3)\otimes U(1)'$ point of view, it is easy to see that there are two massive Dirac ``quarks'' (with $U(1)'$ charge $1/3$ and $2/3$), one massive Dirac ``electron'' (with $U(1)'$ charge 1) and a massless Weyl fermion that we will refer to as the left-handed antineutrino. The $SU(4)\otimes U(1)$ model has an accidental $U(1)_{\ell}$ ($\ell$ for `hidden lepton number') global symmetry (which, not surprising, agrees with the $\{0,0,0,0,s'\}$ nonchiral $U(1)$ discussed above), under which the $\mathbf{4}_{-1}$ and the $\mathbf{\overline{4}}_{-3}$ have the same charge, which is opposite to that of the $\mathbf{6}_2$ and $\mathbf{1}_{-1}$. Note that the vacuum expectation value of $\mathbf{4}^H_{-1}$ does not break the hidden-lepton-number symmetry. 

As far as phenomenology is concerned, in the absence of interactions that connect the hidden sector $SU(4)\otimes U(1)$ model to the SM, the lightest $U(1)'$-charged hidden-sector particle is stable, along with the massless antineutrino. Assuming the hidden $SU(3)$ gauge theory confines in the infrared, the lightest $U(1)'$ particle could be the ``electron'' or an $SU(3)$-neutral baryon-like state. Note that, unlike the SM, there is no separate baryon-number and lepton-number symmetry in this hidden sector. The $W_3$-mediated interactions, for example, connect ``electron'' and ``quark'' states. The presence of massive, stable particles allows one to explore whether there is a viable dark matter candidate in this hidden sector.\footnote{The long-range $U(1)'$ force poses an interesting challenge that may require further model building. There is also the possibility that the dark matter is made up of hidden sector ``atoms''.} We refrain from addressing this in any detail here. 

Yukawa interactions between the SM neutrinos and the hidden sector antineutrinos are forbidden by the SM and the $SU(4)\otimes U(1)$ gauge symmetries. The dimension-five operator, 
\be
{\cal L}_{5} = \frac{(LH)(\mathbf{4}_{-1} (\mathbf{4}^H_{-1})^\dagger)}{\Lambda}+h.c\,,
\label{eq:nus}
\ee
however, after spontaneous symmetry breaking, leads to nonzero neutrino Dirac masses of order $vv_4/\Lambda$, where $v$ and $v_4$ are the vacuum expectation values of the SM Higgs boson and $\mathbf{4}^H_{-1}$, respectively, and $\Lambda$ is the effective messenger scale. Eq.~\eqref{eq:nus} explicitly breaks the $U(1)_{\ell}$ global symmetry and the SM global $U(1)_{B-L}$ symmetry down to a diagonal global lepton number $U(1)_{B-L_{\rm tot}}$, where both the SM leptons and all hidden sector fermions transform. 

The $(3,2)$ models, introduced in Sec.~\ref{sec:U1}, can be interpreted as chiral $SU(3)\otimes SU(2)\otimes U(1)$ models. In particular, the $U(1)$ characterized by $\{a,a,a,d,d\}$ translates into the following $SU(3)\otimes SU(2)\otimes U(1)$ particle content:
\be
({\mathbf{3},\mathbf{2}})_a \,,\quad (\mathbf{\overline{3}},\mathbf{1})_{-a+2d}  \,,\quad (\mathbf{\overline{3}},\mathbf{1})_{-a-2d} \,,\quad (\mathbf{1},\mathbf{2})_{-3a} \,,  \quad (\mathbf{1},\mathbf{1})_{3a+2d} \,, \quad (\mathbf{1},\mathbf{1})_{3a-2d} \,,
\label{model7}
\ee
where $(\mathbf{A},\mathbf{B})_q$ represent a state that transforms as an $A$-plet of $SU(3)$, a $B$-plet under $SU(2)$, and has $U(1)$ charge $q$. Like in the $(5)$ and $(4,1)$ models described above, the $(3,2)$ model is unique as far as the $SU(3)\otimes SU(2)$ quantum numbers are concerned. Like the $(5)$ model, but unlike the $(4,1)$ model, the $SU(3)\otimes SU(2)$ obtained by discarding the $U(1)$ piece of $SU(3)\otimes SU(2)\otimes U(1)$ is also a chiral gauge theory. 

The SM is the $(3,2)$ model characterized by $\{1,1,1,\frac{3}{2},\frac{3}{2}\}$. The $(3,2)$ model can be augmented by another chiral $U(1)$ symmetry characterized by $(a,a,a,d',d')$, where $d'\neq d$. Alternatively, one can choose the non-chiral $U(1)$ symmetry characterized by $(a,a,a,0,0)$ -- in the SM case, this is is  $U(1)_{B-L}$ -- or $(0,0,0,d,d)$ -- in the SM this is a right-handed global $U(1)_R$, where the quark singlets have opposite charge, and the right-handed electron and neutrino also have opposite charge.\footnote{The $U(1)_R$ symmetry is broken by the Yukawa interactions. It is easy to see that the Higgs boson has charge $2d$ and hence does not break $U(1)_{B-L}$.} Combined with hypercharge, the $U(1)_{B-L}$ and $U(1)_R$ charges are not independent. 

Phenomenologically, the $(3,2)$ models provide several interesting opportunities. Since the SM is a $(3,2)$ model, one is allowed to identify some or all of the $SU(3)\otimes SU(2)\otimes U(1)$ with SM simple gauge groups. For example, one can consider adding to the SM a distorted fourth family, where all hypercharge assignments correspond to a different value of $a,d$, i.e., $(a,d)\neq (1,3/2)$. These distorted quarks and leptons would all get masses from the SM Higgs boson, but their electric charges would be different from the SM counterparts. Another possibility is to augment the SM to a gauged, chiral $SU(3)\otimes SU(2) \otimes U(1) \otimes U(1)_{\rm new}$ theory, where the $U(1)_{\rm new}$ charges of the SM particles are characterized by some choice of $(a,d)\neq (1,3/2)$. This is a generalized version of gauging the $U(1)_{B-L}$ symmetry. 

The $(2,2,1)$ models can be interpreted as $SU(2)^2\otimes U(1)$ chiral gauge theories. In particular, the $U(1)$ characterized by $\{a,a,c,c,e\}$ translates into the following $SU(2)^2\otimes U(1)$ particle content:
\begin{eqnarray}
& (\mathbf{1},\mathbf{2})_{-2a-e} \, , \quad (\mathbf{1},\mathbf{2})_{2a-e} \, , \quad (\mathbf{2},\mathbf{1})_{-2c-e} \, , \quad (\mathbf{2},\mathbf{1})_{2c-e} \, , \quad (\mathbf{2},\mathbf{2})_{e} \,, \nn \\
&  (\mathbf{1},\mathbf{1})_{-2a-2c+e} \, , \quad (\mathbf{1},\mathbf{1})_{-2a+2c+e} \, , \quad (\mathbf{1},\mathbf{1})_{2a-2c+e} \, , \quad (\mathbf{1},\mathbf{1})_{2a+2c+e} \,.
\end{eqnarray}
These models can be maximally extended to $SU(2)^2\otimes U(1)^3$ chiral gauge theories by choosing three linearly independent sets of $(a,c,e)$, one for each $U(1)$.

The $(3,1,1)$ and $(2,1,1,1)$ models can be interpreted as $SU(3)\otimes U(1)$ and $SU(2)\otimes U(1)$ chiral gauge theories, respectively. They can be maximally extended, within the $SO(10)$ framework, to $SU(3)\otimes U(1)^2$ and $SU(2)\otimes U(1)^3$. The $(3,1,1)$ models are of the form $\{a,a,a,d,e\}$. For generic values of $(a,d,e)$ the particle content is unique: two $\mathbf{3}$s, two $\mathbf{\overline{3}}$s, all with potentially different $U(1)$ charges, and two singlets, one of which has the highest charge, $3a+d+e$. While the $SU(3)\otimes U(1)$ is chiral as long as none of $a,d,e$ vanish, the $SU(3)$ theory, which one can define by turning off all $U(1)$ charges, is vector-like. it is possible to choose $(a,d,e)$ such that a $\mathbf{3}, \mathbf{\overline3}$ pair has opposite charge or such that one of the singlets has zero charge. In these scenarios, the particle content is smaller. The (2,1,1,1) models are of the form $\{a,a,c,d,e\}$. For generic values of $(a,c,d,e)$ the particle content is unique: four $\mathbf{2}$s, and eight singlets, all with potentially different $U(1)$ charges. One of which has the highest charge, $2a+c+d+e$. For specific choices of $(a,c,d,e)$ charge-zero or vector-like pairs of $\mathbf{2}$s or singlets might emerge, and the chiral particle content might be smaller. 
  
\setcounter{equation}{0}
\section{Extensions}
\label{sec:extensions}

In the previous sections, we concentrated our discussions on chiral gauge theories seeded by $SO(10)$ and the $\mathbf{16}$ representation. In this section, we briefly discuss higher complex representations of $SO(10)$ and other anomaly-free simple gauge theories, including $SO(14)$ and $E_6$. As we will illustrate with a few concrete example, higher representations are required if one is interested in models with more that sixteen fermionic degrees of freedom. Furthermore, larger gauge groups allow one to consider chiral gauge theories with rank larger than five. 

Another reason for pursuing larger groups and higher representations is that the $\mathbf{16}$ of $SO(10)$ might be too restrictive a starting point. In Section \ref{sec:applications}, only a few representations of $SU(N)$ were found when generating the different chiral gauge theories. In particular, in all of Sec.~\ref{sec:applications}, only the fundamental representations\footnote{These are the totally antisimmetric representations, corresponding to the single-column Young tableaux. They include the $\mathbf{10}$, $\mathbf{\overline{10}}$,  $\mathbf{5}$, and $\mathbf{\overline{5}}$ of $SU(5)$, the $\mathbf{6}$, $\mathbf{4}$, $\mathbf{\overline{4}}$ of $SU(4)$, the $\mathbf{3}$ and $\mathbf{\overline{3}}$ of $SU(3)$, and the $\mathbf{2}$ of $SU(2)$.} of $SU(N)$ appear. This is not an accident, but a direct consequence of the fact that we started with the fundamental spinor representation of $SO(10)$.

\subsection{Higher Representations of $SO(10)$}
\label{subsec:highreps}

For higher complex representations of $SO(10)$, we need an expression for generic elements of the $SO(10)$ Cartan subalgebra, similar to Eq.~\eqref{eig-list}. For any complex representation $R$ of $SO(10)$, these will still be functions of five parameters and can also be thought of as vectors on a five-dimensional vector space, $\{a,b,c,d,e\}_R$.  The reason for this, of course, is that $SO(10)$ has rank five. Note that, for any complex representation of $SO(10)$, we can still choose all $a,b,c,d,e\ge 0$, i.e., Eq.~\eqref{eq:positive} is representation independent. 

Starting from the $\mathbf{126}$ representation of $SO(10)$, we find that the particle content of the $SU(5)\otimes U(1)$ model corresponding to $\{1,1,1,1,1\}_{126}$, the (5) model, is, 
\be
\mathbf{1}_{5} \,,\quad \mathbf{\overline{5}}_1 \,,\quad \mathbf{10}_3 \,,\quad \mathbf{\overline{15}}_{-3} \,,\quad \mathbf{\overline{45}}_{-1}  \,,\quad \mathbf{50}_1  \,.
\label{model126_5}
\ee
Here, representations of $SU(5)$ other than the $\mathbf{\overline{5}}$ and $\mathbf{10}$ appear, unlike the $(5)$ associated to the $\mathbf{16}$ of $SO(10)$, Eq.~\eqref{model5}. Like the $(5)$ model associated to the $\mathbf{16}$ of $SO(10)$, the simple $SU(5)$ theory, which one can obtain by simply ignoring the existence of the $U(1)$ group, is also chiral.

On the other hand, starting from the $\mathbf{144}$ representation of $SO(10)$, $\{1,1,1,1,1\}_{144}$ corresponds to
\be
\mathbf{\overline{5}}_7  \,,\quad \mathbf{\overline{45}}_3  \,,\quad \mathbf{5}_3  \,,\quad \mathbf{40}_{-1}  \,,\quad \mathbf{\overline{15}}_{-1} \, , \quad \mathbf{\overline{10}}_{-1}  \,,\quad  \mathbf{24}_{-5}  \,.
\label{model144_5}
\ee
This scenario has some qualitative differences from the $\mathbf{16}$- and $\mathbf{126}$-seeded models, Eq.~\eqref{model5} and Eq.~\eqref{model126_5}, respectively. Eq.~\eqref{model144_5} has no $SU(5)$ singlets and, hence, the highest $U(1)$ charge is not unique. In this case, the highest $U(1)$ charge is 7 and is five-times degenerate (and the five fermions combine into a $\mathbf{\overline{5}}$ of $SU(5)$).

Some technical comments are in order. By only investigating the degeneracy of $U(1)$ charges, it is not possible to distinguish a representation from its complex conjugate. It is also not possible to determine, from the $U(1)$ charges alone, whether a set of degenerate charges corresponds to one or several representations of the nonabelian group. This is the case in Eq.~\eqref{model144_5}, where the $\mathbf{\overline{45}}$ and the $\mathbf{5}$ have the same $U(1)$ charge, and so do the $\mathbf{40}$, the  $\mathbf{\overline{15}}$ and the $\mathbf{\overline{10}}$. In order to extract all this information, one needs to investigate in detail how the different representations of $SO(10)$ organize themselves in terms of representations of the $SU(N)$ nonabelian subgroup of interest. In more detail, we make our assignments by examining how the weight spaces of these representations decompose when $SO(10)$ is broken. For computing the weights of different representations associated to different gauge groups, we made ample use of Ref.~\cite{website}. 


The (4,1) $SU(4)\otimes U(1)$ model corresponding to $\{1,1,1,1,r\}_{126}$ has the following particle content. In what follows, the $\mathbf{4}$s and $\mathbf{\overline{4}}$s, and the  $\mathbf{20}$s and $\mathbf{\overline{20}}$ form vector-like pairs and hence do not contribute to the anomaly cancelation conditions, and could be safely ignored. 
\begin{eqnarray}
& \mathbf{\overline{4}}_{-6} \,, \quad \mathbf{\overline{20}}_{-2} \,, \quad \mathbf{4}_{-2} \,, \quad \mathbf{20}_{2} \,, \quad  \mathbf{\overline{4}}_2 \,, \quad \mathbf{4}_6 \,, \quad  \mathbf{\overline{10}}_{-4-2r} \,, \quad \mathbf{10}_{4-2r} \,, \nn \\  
& \mathbf{15}_{-2r} \,, \quad  \mathbf{20'}_{2r} \,, \quad \mathbf{1}_{2r} \,, \quad \mathbf{1}_{-8+2r} \,, \quad \mathbf{6}_{-4+2r} \, \quad \mathbf{6}_{4+2r} \, \quad \mathbf{1}_{8+2r} \,.
\end{eqnarray}
Similar to the $(5)$ model, here we also obtain larger representations of $SU(4)$. Nonetheless, as in the $\mathbf{16}$ case, Eq.~\eqref{model6}, the $SU(4)$-only model is not chiral, keeping in mind that the $\mathbf{6}$, $\mathbf{15}$ and $\mathbf{20'}$ representations are real. We also examined the $SU(4)\otimes U(1)$ associated to $\{1,1,1,1,r\}_{144}$. There, the $SU(4)$-only particle content is also not chiral.

\subsection{$SO(14)$}
\label{subsec:fourteen}

$SO(14)$ has rank 7, and its smallest complex representation is the fundamental spinor $\mathbf{64}$. All results and intuition developed for and from the $\mathbf{16}$ of $SO(10)$ apply. Here, the elements of the Cartan subalgebra live in a seven-dimensional vector space that can be labelled by $\{a,b,c,d,e,f,g\}$. Chiral models correspond to all 
$a,b,c,d,e,f,g\neq 0$ and degeneracies among $a,b,c,d,e,f,g$ allow one to consider nonabelian chiral gauge theories. 

Starting from the $\mathbf{64}$ of $SO(14)$, one can explore a larger space of models. The following models are too large to fit into $SO(10)$ but can be accommodated by $SO(14)$: $(7), (6,1), (5,2), (4,3),(4,2,1),(3,2,2)$, etc. For example, the (7) model corresponding to $\{1,1,1,1,1,1,1\}$ allows one to define a chiral $SU(7)\otimes U(1)$ model with particle content 
\be
\mathbf{1}_7 \,,\quad \mathbf{\overline{7}}_{-5} \,,\quad \mathbf{21}_3 \,,\quad \mathbf{\overline{35}}_{-1}  \,.
\label{model14_7}
\ee
As is the case of the $\mathbf{16}$ of $SO(10)$, the highest charge is a singlet and only fundamental representations of $SU(7)$ appear. On the other hand, the (6,1) model corresponding to $\{1,1,1,1,1,1,r\}$ can be mapped into an $SU(6)\otimes U(1)$ model 
with particle content 
\be
\mathbf{1}_{6+r} \,,\quad \mathbf{1}_{r-6} \,,\quad \mathbf{6}_{4-r} \,,\quad \mathbf{\overline{6}}_{-4-r} \,,\quad \mathbf{15}_{2+r} \,,\quad \mathbf{\overline{15}}_{r-2} \,,\quad \mathbf{20}_{-r} \,.  
\label{model14_6}
\ee
The $SU(6)$ part of the model is not chiral (the $\mathbf{20}$ representation of $SU(6)$ is real). 

Finally, we discuss a couple of larger imperfect mirrors of the SM that mimic the $SU(M)\otimes SU(N)$ SM structure, $M\neq N\ge 2$ structure of the SM. These include the $\{1,1,1,1,r,r,r\}$ $SU(4)\otimes SU(3)\otimes U(1)$ chiral gauge theory, which has particle content
\begin{eqnarray}
&(\mathbf{1},\mathbf{1})_{4+3r}\,,\quad (\mathbf{4},\mathbf{3})_{2+r} \,,\quad (\mathbf{1},\mathbf{\overline{3}})_{4-r}\,,\quad (\mathbf{6}, \mathbf{1})_{3r}\,,\quad (\mathbf{4}, \mathbf{1})_{2-3r}\,, \nn \\
&(\mathbf{6},\mathbf{\overline{3}})_{-r}\,,\quad (\mathbf{\overline{4}},\mathbf{3})_{r-2}\,,\quad (\mathbf{\overline{4}},\mathbf{1})_{-2-3r}\,,\quad (\mathbf{1},\mathbf{1})_{3r-4}\,,\quad (\mathbf{1},\mathbf{\overline{3}})_{-4-r}.
\end{eqnarray}
The $SU(4)\otimes SU(3)$ model is also chiral. 

On the other hand, the $\{1,1,1,1,1,r,r\}$ $SU(5) \otimes SU(2)\otimes U(1)$ chiral gauge theory has particle content
\begin{eqnarray}
&(\mathbf{1},\mathbf{1})_{5+2r}\,,\quad (\mathbf{5}, \mathbf{2})_{3}\,,\quad (\mathbf{1},\mathbf{1})_{5-2r}\,,\quad (\mathbf{10},\mathbf{1})_{1+2r}\,,\nn \\
&(\mathbf{10},\mathbf{1})_{1-2r}\,,\quad (\mathbf{\overline{10}},\mathbf{2})_{-1}\,,\quad (\mathbf{\overline{5}},\mathbf{1})_{-2r-3}\,,\quad (\mathbf{\overline{5}},\mathbf{1})_{2r-3}\,,\quad (\mathbf{1},\mathbf{2})_{-5}.
\end{eqnarray}

\subsection{$E_6$}
\label{subsec:Esix}

The exceptional group $E_6$ has rank 6, and its smallest complex representation is the $\mathbf{27}$ (there are two inequivalent 27-dimensional representations). As in the $SO(10)$ and $SO(14)$ cases, we can represent the elements of the $E_6$ Cartan subalgebra as six-dimensional vectors, labelled $\{a,b,c,d,e,z\}$. Explicitly, the elements of the Cartan subalgebra for one of the $\mathbf{27}$ representations can be written as
\begin{eqnarray}
H(a,b,c,d,e,z) & \propto & {\rm diag}\{a + b + c + d + e + z, a + b + c - d - e + z,  a + b - c + d - e + z,  a + b - c - d + e + z, \nn \\
 & & a - b + c + d - e + z,  a - b + c - d + e + z, -a + b + c + d - e + z, a - b - c + d + e + z, \nn \\
 & & -a + b + c - d + e + z, -a + b - c + d + e + z, a - b - c - d - e + z, -a - b + c + d + e + z, \nn \\ 
 & & -a + b - c - d - e + z, -a - b + c - d - e + z, -a - b - c + d - e + z, -a - b - c - d + e + z, \nn \\
 & & 2 a - 2 z, 2 b - 2 z, 2 c - 2 z, 2 d - 2 z, 2 e - 2 z, -2 e - 2 z, -2 d - 2 z, -2 c - 2 z, -2 b - 2 z, -2 a - 2 z, 4z \nn \}. \\
 & & 
\end{eqnarray}
The first sixteen charges look remarkably similar to the entries in Eq.~\eqref{eig-list}. This is, of course, not an accident; $E_6$ contains $SO(10) \otimes U(1)$ as a maximal subgroup, and the $\mathbf{27}$ decomposes into $\mathbf{16}_1 \oplus \mathbf{10}_{-2} \oplus \mathbf{1}_{4}$, where the subscript indicates the $U(1)$ charge. The first sixteen charges are the $U(1)$ charges of the $\mathbf{16}$ of $SO(10)$ up to an additive factor of $z$. If $z \neq 0$, then the charges of the $\mathbf{16}$, by themselves, do not satisfy the anomaly equations, Eq.~\eqref{anomaly-eqs}. However, when one factors in the eleven remaining charges, the linear and cubic anomalies vanish. In this scenario, the chiral fermions of the $\mathbf{16}$ are necessarily accompanied by vector-like fermions which belong to the $\mathbf{10}$ and $\mathbf{1}$ representations. 

The discussion of the interpretation of these charge assignments follows our discussion in Sec.~\ref{sec:applications}. The coefficients $a,b,c,d,e$ inform us about how states are grouped into representations of nonabelian subgroups, depending on which subsets of coefficients are equal, as before. The number $z$ describes a uniform shift of the $U(1)$ charges within each multiplet of $SO(10)$, while maintaining that the 27 charges, taken together, are nonanomalous. As before, we are free to impose that $a,b,c,d,e$ are all positive without loss of generality. However, $z$ must be allowed to have either sign; we highlight this by showing the charge assignments in the two cases $\{a,b,c,d,e,z\} = \{1,1,1,1,1,+1\}$ and $\{1,1,1,1,1,-1\}$, which both possess $SU(5)\otimes U(1)$ symmetry:
\begin{eqnarray}
\{1,1,1,1,1,+1\}: \quad & \mathbf{1}_6 \, , \quad \mathbf{\overline{5}}_{-2} \, , \quad \mathbf{10}_2 \, , \quad \mathbf{5}_{-4} \, , \quad \mathbf{\overline{5}}_0 \, , \quad \mathbf{1}_4 \,, \nn \\
\{1,1,1,1,1,-1\}: \quad & \mathbf{1}_{4} \, , \quad \mathbf{\overline{5}}_{-4} \, , \quad \mathbf{10}_0 \, , \quad \mathbf{5}_0 \, , \quad \mathbf{\overline{5}}_4 \, , \quad \mathbf{1}_{-4} \,. \nn \\
\end{eqnarray}
Both scenarios are different from one another and different from Eq.~\eqref{model5}.


\setcounter{equation}{0}
\section{Discussion}

Chirality is an intriguing feature of nature. Quantum anomalies render generic chiral gauge theories inconsistent, and yet, in the SM, these are mysteriously absent thanks to a set of delicate cancellations. It seems pertinent to ask the questions: How do these cancellations come to happen? Do anomaly cancellations reveal that there is more structure at very small distance scales?
 

The standard explanation is to posit that the SM is embedded into a complex representation of a higher-dimensional, anomaly-free unification gauge theory, such as $SO(10)$. If $SO(10)$ unification is at play, then anomaly cancellations are naturally explained and the existence of chiral fermions is simple to understand. On top of explaining why anomalies cancel, an $SO(10)$ GUT also predicts the unification of the gauge couplings, the existence of new, ultra-heavy particles, proton decay, etc. In this work we concentrated on a humbler, more agnostic task. If chirality were a fundamental principle, and not assuming unification, we have asked what are other chiral, anomaly-free gauge theories. Inspired by the SM embedding in the $SO(10)$ -- but making no assumptions regarding gauge coupling unification or the existence of a GUT gauge theory -- we have devised a general method to construct such theories from scratch. 

Due to the anomaly cancellation conditions, these imperfect mirrors of the SM are necessarily nonminimal. In Ref.~\cite{firstpaper}, a concrete $U(1)$ chiral gauge theory was explored. There, it was demonstrated how the rich, chiral structure provides the necessary ingredients for simultaneously addressing the dark matter and neutrino mass puzzles. In this paper, we provide an algorithm that makes it possible to characterize and study an infinite number of qualitatively different chiral models, many of which could have interesting phenomenological applications. 

In order to prove the statements and claims in this paper, relatively heavy use of group theory was made. Nonetheless, it should be emphasized that an important result advocated here is that much information about chiral models can be obtained without any group theory analysis whatsoever. In particular, a result of this work is that the particle content of anomaly-free chiral models, including charges and nonabelian representations, can be easily obtained by following the procedure described in Secs.~\ref{sec:U1} and \ref{sec:nonabelian}. A summary of the procedure, for imperfect mirrors based on the \tbf{16} of $SO(10)$, is as follows: (i) Pick the gauge group of the model of the form $\msc{G} = SU(n_1)\otimes \dots SU(n_k)\otimes U(1)^m$ subject to the requirements that the rank of $\msc{G}\le 5$, $n_1+\ldots+n_k\le 5$, and $m\geq 1$; (ii) pick $m$ sets of five numbers $\{a,b,c,d,e\}$ such that they can be divided into sets of $n_1,\dots,n_k$ equal numbers, and such that the $m$ different sets are linearly independent; and (iii) introduce fermions with $U(1)$ charges given by Eq.~\eqref{eig-list}. The fermions will automatically organize into anomaly-free multiplets of $\msc{G}$. For example, the choice $\{1,1,\frac{1}{2},\frac{1}{2},2\}$, $\{\frac{1}{2},\frac{1}{2},\frac{3}{2},\frac{3}{2},1\}$ will yield a set of chiral fermions nontrivially charged under $SU(2)^2\otimes U(1)^2$.

This algorithm, which does not require group theory savviness, works for models based on complex representations of groups of the form $SO(4N+2)$. Models based on the group $E_6$ -- the only other compact, anomaly-free Lie group with complex representations -- can be treated similarly, as briefly outlined in Sec.~\ref{sec:extensions}, although the identification of nonabelian symmetries is not as straightforward.

Some well-known extensions of the Standard Model, even ones that can be obtained by descending from $SO(10)$ unification, cannot be directly obtained from the procedure proposed here. The most prominent example is the Pati-Salam model~\cite{Pati:1974yy} based on the gauge group $\msc{G} = SU(4)\otimes SU(2)_L\otimes SU(2)_R$. The problem is readily apparent:  the ranks of the Pati-Salam group and $SO(10)$ are the same, but the numbers $n_i$ in this case satisfy $n_1 + n_2 + n_3 = 8$. However, a simple generalization of our procedure to account for this case, under investigation, seems to exist.


While it was not the purpose of this work to focus on specific models, we did briefly discuss, as examples of the power of this procedure, chiral models with large gauge groups that can arise out of the \tbf{16} of $SO(10)$. We believe that many of these are interesting in their own right. They include: (i) a unique $SU(5)\otimes U(1)$ model, essentially the Georgi-Glashow $SU(5)$ unification; (ii) A new class of $SU(4)\otimes U(1)^2$ scenarios that have not, as far we as can tell, been explored before and that, in particular, do not correspond to Pati-Salam unification; and (iii) $SU(3)\otimes SU(2)\otimes U(1)$ models that mimic the nonabelian representations of the SM but where the charges with respect to the abelian counterpart of hypercharge are different from those of the SM fermions. Finally, we also discussed, very briefly, models one can construct starting with larger, complex representations of $SO(10)$, or with larger simple gauge groups: $SO(14)$ and $E_6$.

\begin{acknowledgments}
AdG and DH thank Bogdan Dobrescu for inspiring conversations. This work is supported in part by the DOE grant \#DE-FG02-91ER40684.  
\end{acknowledgments}

\end{document}